\documentclass[australian,twocolumn, showpacs, showkeys]{revtex4-1}
\usepackage[T1]{fontenc}
\usepackage[latin9]{inputenc}
\usepackage{units}
\usepackage{amsmath}
\usepackage{graphicx}
\usepackage{esint}
\usepackage{babel}
\begin{document}

\title{Efficient numerical solution of the time fractional diffusion equation
by mapping from its Brownian counterpart}

\author{Peter W. \surname{Stokes}}

\email[Electronic address: ]{peter.stokes@my.jcu.edu.au}

\affiliation{College of Science, Technology \& Engineering, James Cook University,
Townsville, QLD 4811, Australia}

\author{Bronson \surname{Philippa}}

\affiliation{College of Science, Technology \& Engineering, James Cook University,
Townsville, QLD 4811, Australia}

\author{Wayne \surname{Read}}

\affiliation{College of Science, Technology \& Engineering, James Cook University,
Townsville, QLD 4811, Australia}

\author{Ronald D. \surname{White}}

\affiliation{College of Science, Technology \& Engineering, James Cook University,
Townsville, QLD 4811, Australia}
\begin{abstract}
The solution of a Caputo time fractional diffusion equation of order
$0<\alpha<1$ is expressed in terms of the solution of a corresponding
integer order diffusion equation. We demonstrate a linear time mapping
between these solutions that allows for accelerated computation of
the solution of the fractional order problem. In the context of an
$N$-point finite difference time discretisation, the mapping allows
for an improvement in time computational complexity from $O\left(N^{2}\right)$
to $O\left(N^{\alpha}\right)$, given a precomputation of $O\left(N^{1+\alpha}\ln N\right)$.
The mapping is applied successfully to the least squares fitting of
a fractional advection-diffusion model for the current in a time-of-flight
experiment, resulting in a computational speed up in the range of
one to three orders of magnitude for realistic problem sizes.
\end{abstract}



\maketitle

\section{\label{sec:Section-I}Introduction}

Derivatives of non-integer order have been particularly successful
in describing a variety of complex processes with memory effects.
These include applications in statistical finance \citep{scalas2000fractional},
economic modelling \citep{vskovranek2012modeling}, image processing
\citep{Janev2011}, quantum systems \citep{wu2014fractional} and
kinetics \citep{sokolov2002fractional,pagnini2014short,philippa2014generalized,philippa2011analytic,sibatov2007fractional,sagi2012observation,toledo2014fractional}.
This paper will focus on the numerical solution of a fractional kinetics
description of anomalous diffusion. Unlike normal diffusion, whose
mean squared displacement grows linearly with time, the anomalous
diffusion considered here is characterised by a mean squared displacement
that grows sublinearly according to a power law of the form $t^{\alpha}$
with $0<\alpha<1$ \citep{Metzler2002,Sokolov2005,sagi2012observation,metzler1999anomalous,metzler2000random}.
A number of stochastic approaches are capable of describing this kind
of anomalous diffusion \citep{metzler1999anomalous,metzler2000random,klafter1987stochastic,montroll1965random,Angstmann2014,hilfer1995fractional,srokowski2014anomalous}.
For example, Scher and Montroll \citep{scher1975anomalous} used a
continuous time random walk (CTRW) model to describe the anomalous
transport of charge carriers in disordered semiconductors. In this
case, anomalous behaviour arises due to the localised trapping of
charge carriers. To describe this trapping, a CTRW was chosen that
sampled from a distribution of trapping times of the power law form
$w\left(t\right)\sim t^{-\left(1+\alpha\right)}$. Here, $\alpha$
describes the severity of the trapping, with smaller values of $\alpha$
corresponding to increasingly severe traps. In disordered semiconductors,
$\alpha$ arises physically from the energetic width of the density
of localised states \citep{philippa2014generalized,sibatov2007fractional,Bisquert2003}.
It has been rigorously shown \citep{barkai2000continuous,metzler2000random,metzler2004restaurant}
that a CTRW of this form can be described by a diffusion equation
with a time derivative of fractional order $\alpha$. In this paper,
we are concerned with the numerical solution of a Caputo fractional
advection-diffusion model for the current in a time-of-flight experiment
for a disordered semiconductor \citep{philippa2011analytic,uchaikin2008fractional,sibatov2007fractional,Sandev2011,Mainardi1996}
\begin{equation}
_{0}^{\mathrm{C}}\mathcal{D}_{t}^{\alpha}u\left(t,x\right)=D_{\mathrm{L}}\frac{\partial^{2}}{\partial x^{2}}u\left(t,x\right)-W\frac{\partial}{\partial x}u\left(t,x\right),\label{eq:mainFDE}
\end{equation}
where $W$ is a generalised drift velocity, $D_{\mathrm{L}}$ is a
generalised diffusion coefficient and the operator for Caputo fractional
differentiation of order $0<\alpha<1$ is defined in terms of the
convolution integral \citep{caputo1967linear} 
\begin{equation}
_{0}^{\mathrm{C}}\mathcal{D}_{t}^{\alpha}f\left(t\right)\equiv\frac{1}{\Gamma\left(1-\alpha\right)}\int_{0}^{t}\mathrm{d}\tau\left(t-\tau\right)^{-\alpha}f^{\prime}\left(\tau\right).\label{eq:caputoDef}
\end{equation}
Note that the normal advection-diffusion equation can be recovered
in the relevant limit of no trapping
\begin{equation}
\lim_{\alpha\rightarrow1}{}_{0}^{\mathrm{C}}\mathcal{D}_{t}^{\alpha}u\left(t,x\right)=\frac{\partial}{\partial t}u\left(t,x\right).
\end{equation}

Numerous methods exist \citep{Gao2011586,gao2014new,zhang2014finite,yang2014orthogonal,bhrawy2014spectral,fu2013boundary,wu2010fractional}
for finding the numerical solution of fractional differential equations
of the form of Eq. (\ref{eq:mainFDE}). Many of these are direct analogues
to approaches that are also applicable to integer order differential
equations. This is to be expected with the definition of fractional
differentiation (\ref{eq:caputoDef}) defined in terms of both differentiation
and integration. Unfortunately, when solving fractional differential
equations numerically there is an increase \citep{podlubny1998fractional}
in time computational complexity over that encountered when solving
differential equations of integer order. This is due to the global
nature of fractional differentiation and, as in the case of anomalous
diffusion, can be interpreted as a result of the system having memory.
Consequently, any numerical algorithm that computes the solution at
a present point in time requires the entire solution history to do
so. In the context of an $N$-point finite difference time discretisation,
this causes a time computational complexity increase from $O\left(N\right)$
to $O\left(N^{2}\right)$ \citep{Ford2001}.

A number of approaches have been proposed to accelerate the computation
of the numerical solution of fractional differential equations \citep{podlubny1998fractional,Ford2001,fukunaga2011high,Diethelm2011,Gong2014}.
As this added computational complexity stems from the memory inherent
to the system, many of these approaches involve restricting this memory
in some way. Podlubny \citep{podlubny1998fractional} considered this
approach by introducing the \textit{fixed memory principle}, which
amounts to truncating the convolution integral in the definition of
fractional differentiation (\ref{eq:caputoDef}). In effect, this
restricts the memory of the system to a fixed interval of time into
the past, subsequently allowing for the solution to be found numerically
in $O\left(N\right)$ in exchange for some loss in solution accuracy.
Unfortunately, the only way to guarantee the accuracy of a numerical
method used in conjunction with the fixed memory principle is to choose
a fixed interval of time that encompasses the entire history of the
solution, returning the computational complexity to $O\left(N^{2}\right)$.
Ford and Simpson \citep{Ford2001} demonstrated exactly this and,
as an alternative, introduced the \textit{logarithmic memory principle},
which samples from the solution history in a logarithmic fashion,
allowing for the solution to be found in $O\left(N\ln N\right)$ without
compromise in solution accuracy. Finally, a number of parallel computing
algorithms have also been introduced \citep{Diethelm2011,Gong2014}.
These approaches are viable ways for accelerating the computation
of the solution although, as they often involve splitting the problem
into smaller problems of the same computational complexity, they are
ultimately still of $O\left(N^{2}\right)$.

In Section \ref{sec:Section-II} of the current study, we show that
the solution to the fractional advection-diffusion equation (\ref{eq:mainFDE})
can be related to the solution of the normal advection-diffusion equation
through a linear mapping in time. This mapping relationship, which
takes the form of a matrix multiplication, provides an approach for
the numerical acceleration of the fractional solution. In Section
\ref{sec:Section-III}, an algorithm for the computation of the matrix
that defines the linear mapping is presented that utilises the fast
Fourier transform. Additionally, we show that many elements of this
matrix may contribute negligibly to the solution and so can be neglected,
subsequently allowing for even further acceleration. In Section \ref{sec:Section-IV},
we demonstrate the accuracy of this mapping approach by benchmarking
the numerical solution of a fractional relaxation equation against
its exact analytic solution. In Section \ref{sec:Section-V}, this
mapping is applied successfully to accelerate the fitting of Eq. (\ref{eq:mainFDE})
to experimental data for a time-of-flight experiment. Finally, in
Section \ref{sec:Section-VI}, we present conclusions and briefly
list possible applications of our approach to various generalisations
of the considered fractional-order problem.

\section{\label{sec:Section-II}Mapping between normal and fractional diffusion}

In this section, we will explore accelerating the numerical solution
of the fractional advection-diffusion equation (\ref{eq:mainFDE})
by relating it to the solution of the normal advection-diffusion equation
\begin{equation}
\frac{\partial}{\partial\tau}v\left(\tau,x\right)=D_{\mathrm{L}}\frac{\partial^{2}}{\partial x^{2}}v\left(\tau,x\right)-W\frac{\partial}{\partial x}v\left(\tau,x\right),\label{eq:mainIDE}
\end{equation}
where $\tau$ has fractional units of time due to the presence of
the generalised transport coefficients $D_{\mathrm{L}}$ and $W$.
By enforcing both equivalent initial conditions and boundary conditions,
we can relate these solutions using the known integral transform relationship
\citep{bouchaud1990anomalous,klafter1994probability,saichev1997fractional,barkai2000fractional,barkai2001fractional}
\begin{equation}
u\left(t,x\right)=\int_{0}^{\infty}\mathrm{d}\tau A\left(\tau,t\right)v\left(\tau,x\right),\label{eq:continuousMap}
\end{equation}
which also holds true for any other shared \textit{linear }\emph{spatial}
operator in the considered advection-diffusion equations. Here, the
kernel is defined 
\begin{equation}
A\left(\tau,t\right)\equiv\mathcal{L}^{-1}\left\{ s^{\alpha-1}\mathrm{e}^{-s^{\alpha}\tau}\right\} =\frac{\partial}{\partial\tau}\left[1-L_{\alpha}\left(\frac{t}{\sqrt[\alpha]{\tau}}\right)\right],\label{eq:kernelDef}
\end{equation}
where $\mathcal{L}$ denotes the Laplace transform and $L_{\alpha}\left(t\right)$
is the one-sided Lévy distribution, which is expressible in terms
of the one-sided Lévy density $l_{\alpha}\left(t\right)$ as 
\begin{equation}
L_{\alpha}\left(t\right)\equiv\int_{0}^{t}\mathrm{d}\tau l_{\alpha}\left(\tau\right),\quad\mathcal{L}l_{\alpha}\left(t\right)\equiv\mathrm{e}^{-s^{\alpha}}.
\end{equation}
This integral relationship is known as a subordination transformation,
where $A\left(\tau,t\right)$ is the probability distribution function
providing subordination of the random process governed by Eq. (\ref{eq:mainFDE})
on the physical time scale $t$ to that governed by Eq. (\ref{eq:mainIDE})
on the operational time scale $\tau$ \citep{bochner2012harmonic}.

In order to determine the fractional order solution numerically, we
wish to find a discrete analogue of this transform. We note that this
relationship acts on time alone, independent of space. As such, in
what follows, we shall consider the solutions $u\left(t,x\right)$
and $v\left(\tau,x\right)$ solely as functions of time and reintroduce
spatial dependence at a later point. Performing separation of variables,
we can instead consider the ordinary time differential equations
\begin{eqnarray}
_{0}^{\mathrm{C}}\mathcal{D}_{t}^{\alpha}u\left(t\right) & = & \lambda u\left(t\right),\label{eq:fracODE}\\
\frac{\mathrm{d}}{\mathrm{d}\tau}v\left(\tau\right) & = & \lambda v\left(\tau\right).\label{eq:intODE}
\end{eqnarray}
where $\lambda$ is the separation constant or eigenvalue of the shared
spatial operator. We will now perform a finite difference time discretisation
of these ordinary differential equations. We will denote time steps
by superscripts $u^{n}\equiv u\left(n\Delta t\right)$, where $\Delta t$
is the time step size and $n=0,\dots,N$ is the time step number with
$N$ being the total number of time steps and $t\equiv N\Delta t$
being the present point in time. To numerically approximate the fractional
time derivative we will make use of the L1 algorithm \citep{oldham1974fractional},
which was introduced by Oldham and Spanier to approximate the Riemann-Liouville
fractional derivative. This algorithm has since been applied by a
number of authors \citep{Gao2011586,gao2014new,Lin20071533,Murio2008,Zhao20116061}
to the Caputo fractional derivative, resulting in the approximation
\begin{equation}
_{0}^{\mathrm{C}}\mathcal{D}_{t}^{\alpha}u\left(t\right)=\Delta t^{-\alpha}\sum_{n=1}^{N}w_{n}\left(u^{N-n+1}-u^{N-n}\right)+O\left(\Delta t\right),\label{eq:fracDiscrete}
\end{equation}
where we have the quadrature weights defined
\begin{equation}
w_{n}\equiv\frac{n^{1-\alpha}-\left(n-1\right)^{1-\alpha}}{\Gamma\left(2-\alpha\right)}.
\end{equation}
This discretisation of the Caputo fractional derivative includes the
limiting case where $\alpha\rightarrow1$ from which we can recover
the Euler method
\begin{equation}
\frac{\mathrm{d}}{\mathrm{d}\tau}v\left(\tau\right)=\frac{v^{N}-v^{N-1}}{\Delta\tau}+O\left(\Delta\tau\right).\label{eq:intDiscrete}
\end{equation}
Applying these discretisations to the ordinary differential equations,
respectively (\ref{eq:fracODE}) and (\ref{eq:intODE}), yields the
recurrence relationships for the finite difference solution approximations
\begin{eqnarray}
\left(1-\lambda\frac{\Delta t^{\alpha}}{w_{1}}\right)u^{N} & = & \hat{w}_{N}u^{0}+\sum_{n=1}^{N-1}\left(\hat{w}_{n}-\hat{w}_{n+1}\right)u^{N-n},\quad\quad\label{eq:fracRecurrence}\\
\left(1-\lambda\Delta\tau\right)v^{N} & = & v^{N-1},\label{eq:intRecurrence}
\end{eqnarray}
where we have introduced the normalised quadrature weights $\hat{w}_{n}\equiv w_{n}/w_{1}$.
As expected, the fractional order solution at each time step depends
on the entire solution history, while the integer order solution depends
only the nearest prior point in the neighbourhood of the present.
We can solve these recurrence relationships analytically for the present
time step in terms of their respective initial conditions
\begin{eqnarray}
u^{N} & = & \sum_{n=1}^{N}a_{Nn}\frac{u^{0}}{\left(1-\lambda\frac{\Delta t^{\alpha}}{w_{1}}\right)^{n}},\label{eq:fracSolution}\\
v^{N} & = & \frac{v^{0}}{\left(1-\lambda\Delta\tau\right)^{N}},\label{eq:intSolution}
\end{eqnarray}
where $a_{Nn}$, which is yet to be determined, denotes the $n$-th
weight in the weighted sum for the fractional order solution at the
$N$-th time step. If we choose the integer order initial condition
to coincide with the fractional one $v^{0}=u^{0}$ and also choose
a time step size for the integer order case of $\Delta\tau=\Delta t^{\alpha}/w_{1}$
we can relate the solution to the fractional order problem directly
to the solution of the integer order one as
\begin{equation}
u^{N}=\sum_{n=1}^{N}a_{Nn}v^{n}.\label{eq:singleMap}
\end{equation}
This is a discrete analogue of the continuous integral relationship
(\ref{eq:continuousMap}) and so the weights $a_{Nn}$ can be interpreted
as quadrature weights. We should expect this discrete analogue to
coincide with the continuous relationship in the limit of many time
steps $N$. Most generally, reintroducing spatial dependence and considering
all time steps, we can write each weighted sum in the form of Eq.
(\ref{eq:singleMap}) using the matrix multiplication
\begin{equation}
\mathbf{U}=\mathbf{A}\mathbf{V},\label{eq:fullMap}
\end{equation}
where we have the matrix of quadrature weights 
\begin{equation}
\mathbf{A}=\left[\begin{array}{ccc}
a_{11} & 0 & 0\\
\vdots & \ddots & 0\\
a_{N1} & \cdots & a_{NN}
\end{array}\right],
\end{equation}
which allows for mapping from the integer order solution matrix 
\begin{equation}
\mathbf{V}=\left[\begin{array}{ccc}
\text{---} & \mathbf{v}^{1} & \text{---}\\
 & \vdots\\
\text{---} & \mathbf{v}^{N} & \text{---}
\end{array}\right],
\end{equation}
to the fractional order solution matrix
\begin{equation}
\mathbf{U}=\left[\begin{array}{ccc}
\text{---} & \mathbf{u}^{1} & \text{---}\\
 & \vdots\\
\text{---} & \mathbf{u}^{N} & \text{---}
\end{array}\right],
\end{equation}
where the rows of these solution matrices correspond to the spatial
solution at each time step for the same spatial points. As the mapping
matrix $\mathbf{A}$ is lower triangular, determining the solution
matrix $\mathbf{U}$ using this matrix multiplication is of $O\left(N^{2}\right)$.
This is no better than directly applying Eq. (\ref{eq:fracRecurrence})
to find the solution recursively. Fortunately, this is only the case
if we absolutely require the solution at \emph{every} time step. Indeed,
if we are content with the solution at a subset of the overall time
steps, we can perform the matrix multiplication in Eq. (\ref{eq:fullMap})
partially in $O\left(N\right)$. Consider, for example, stability
limitations such as the Courant-Friedrichs-Lewy condition \citep{thomas1995numerical}
that arise in explicit finite difference schemes and may require time
steps smaller than would otherwise be needed. In such a situation,
we can solve the integer order problem with sufficiently small time
steps (to satisfy the stability criterion), and then map it onto the
fractional problem using sparser time steps. Of course, the usefulness
of this approach also depends on the computational complexity in computing
the required rows of the mapping matrix. Fortunately, as the solution
mapping depends solely on the operator of fractional differentiation,
the mapping matrix can be precomputed for a given value of $\alpha$
and used repeatedly. The precise computational complexity for computing
the mapping matrix will be considered in Section \ref{sec:Section-III}.

\section{\label{sec:Section-III}The solution mapping matrix}

In this section, we address the problem of efficiently computing and
applying the mapping matrix $\mathbf{A}$, present in Eq. (\ref{eq:fullMap})
for the numerical relationship between integer and fractional order
solutions.

\subsection{Computation of the mapping matrix $\mathbf{A}$ using the fast Fourier
transform}

Substitution of the fractional finite difference solution approximation
(\ref{eq:fracSolution}) back into its recurrence relationship (\ref{eq:fracRecurrence})
allows us to express the elements of the mapping matrix $\mathbf{A}$
in the form of a generating function recurrence relationship
\begin{equation}
A_{n}\left(x\right)=\Omega\left(x\right)A_{n-1}\left(x\right),\label{eq:funRecurrence}
\end{equation}
where we have the generating function for the $n$-th column of the
mapping matrix
\begin{equation}
A_{n}\left(x\right)\equiv\sum_{m\geq1}a_{mn}x^{m},
\end{equation}
with the first column given by the initial condition weights from
Eq. (\ref{eq:fracRecurrence})
\begin{equation}
A_{1}\left(x\right)\equiv\sum_{m\geq1}\hat{w}_{m}x^{m},
\end{equation}
and the generating function of past time step weights from Eq. (\ref{eq:fracRecurrence})
\begin{equation}
\Omega\left(x\right)\equiv\sum_{m\geq1}\left(\hat{w}_{m}-\hat{w}_{m+1}\right)x^{m}.
\end{equation}
The Cauchy product \citep{Apostol1997} allows us to write this generating
function recurrence relationship explicitly using a discrete linear
convolution
\begin{equation}
\left[\begin{array}{c}
a_{nn}\\
\vdots\\
a_{Nn}
\end{array}\right]=\left[\begin{array}{c}
\hat{w}_{1}-\hat{w}_{2}\\
\vdots\\
\hat{w}_{N-n+1}-\hat{w}_{N-n+2}
\end{array}\right]\ast\left[\begin{array}{c}
a_{n-1,n-1}\\
\vdots\\
a_{N-1,n-1}
\end{array}\right],\label{eq:convRecurrence}
\end{equation}
where the initial column vector is provided by its generating function
$A_{1}\left(x\right)$
\begin{equation}
\left[\begin{array}{c}
a_{11}\\
\vdots\\
a_{N1}
\end{array}\right]=\left[\begin{array}{c}
\hat{w}_{1}\\
\vdots\\
\hat{w}_{N}
\end{array}\right].
\end{equation}
This convolution representation can be implemented using the fast
Fourier transform, allowing for the computation of an $N\times N$
mapping matrix in $O\left(N^{2}\ln N\right)$. Evidently, determining
the mapping matrix alone is more computationally intensive than finding
the finite difference solution recursively in only $O\left(N^{2}\right)$.
Certain situations exist, however, where the mapping matrix may be
precomputed and reused, allowing for computational benefit even with
this larger computational complexity. One such situation is the focus
of Section \ref{sec:Section-V}, where the least squares fit of a
fractional order model to experimental data is considered. Fortunately,
as described in the following subsection, we are not limited to only
these situations when it comes to useful application of this solution
mapping.

\subsection{Column truncation of the mapping matrix $\mathbf{A}$}

\begin{figure}
\includegraphics[scale=0.2]{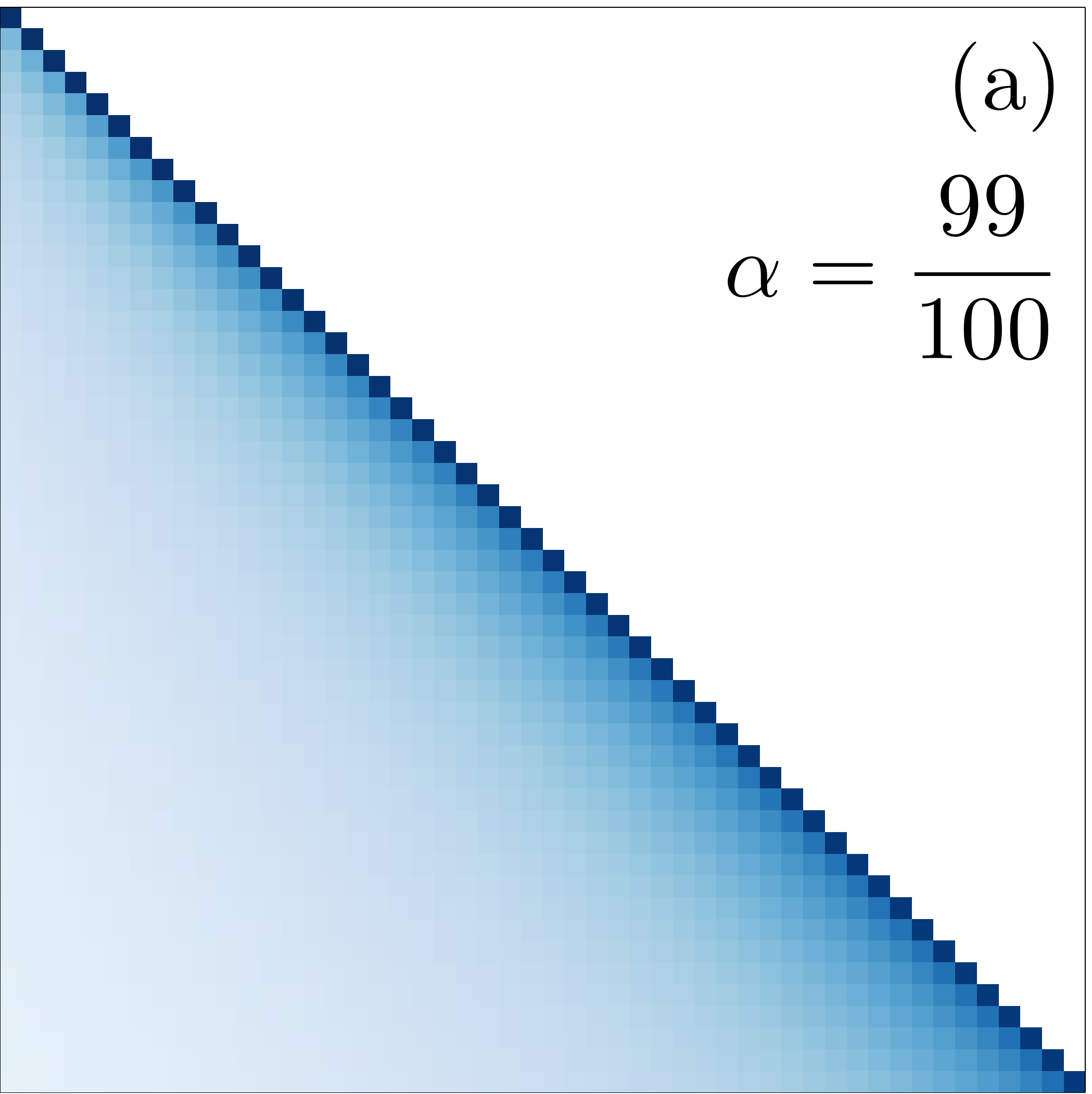}$\,$\includegraphics[scale=0.2]{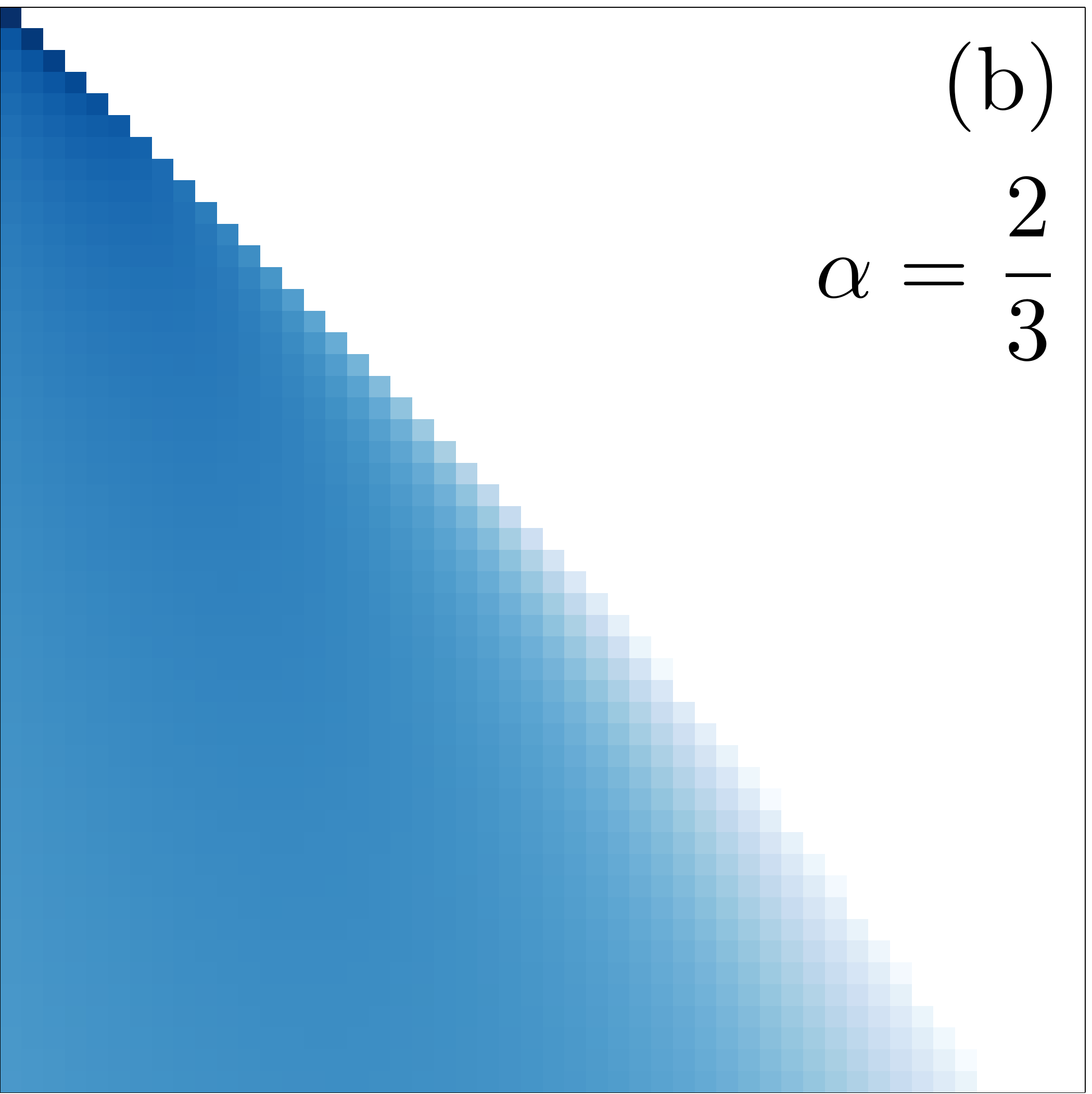}

\includegraphics[scale=0.2]{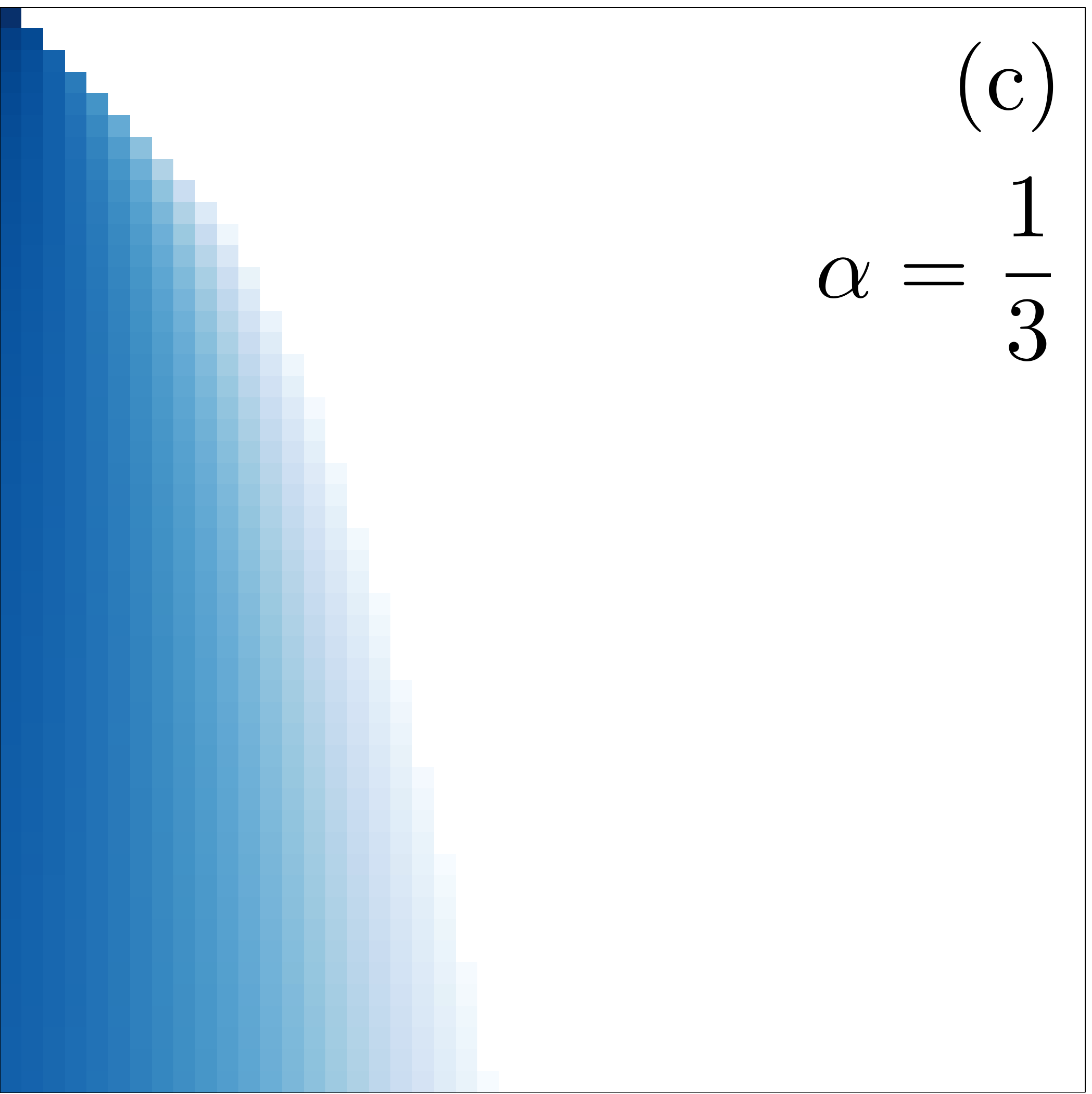}$\,$\includegraphics[scale=0.2]{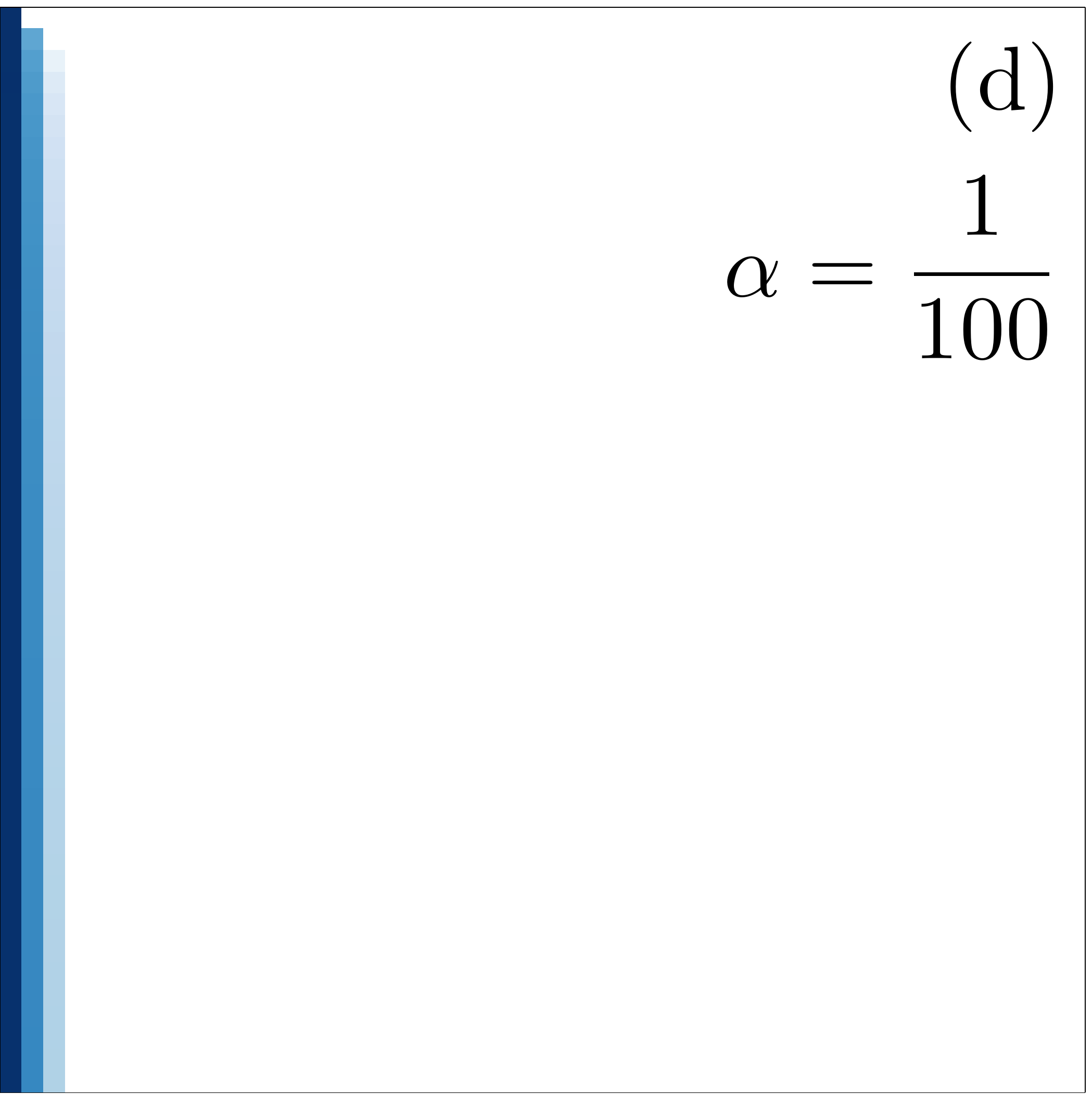}

\includegraphics[scale=0.705]{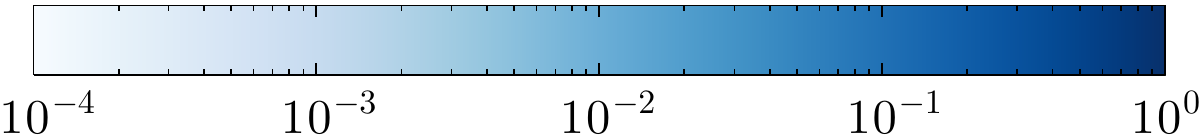}

\protect\caption{\label{fig:truncationIllustration}Illustration of the matrix $\mathbf{A}$
that maps from the solution of the normal diffusion equation (\ref{eq:mainIDE})
to the solution of the order $\alpha$ fractional diffusion equation
(\ref{eq:mainFDE}). Each matrix is of size $50\times50$ with elements
that have been coloured according to their magnitude on a logarithmic
scale. (a) As $\alpha\rightarrow1$, the identity matrix is recovered,
corresponding to the fractional and integer order solutions coinciding.
(b-c) As $\alpha$ decreases, the matrix is dominated by elements
with a lower column number, indicating that the early time solution
to the integer order problem becomes increasingly significant. (d)
As $\alpha\rightarrow0$, the matrix approaches having only an initial
column of ones, which corresponds to a time-invariant solution. This
rapid decrease in element magnitude suggests the possibility of column-wise
truncation of the mapping matrix, allowing for improved efficiency
in both its computation and application, especially for small values
of $\alpha$.}
\end{figure}

\begin{figure}
\includegraphics[scale=0.85]{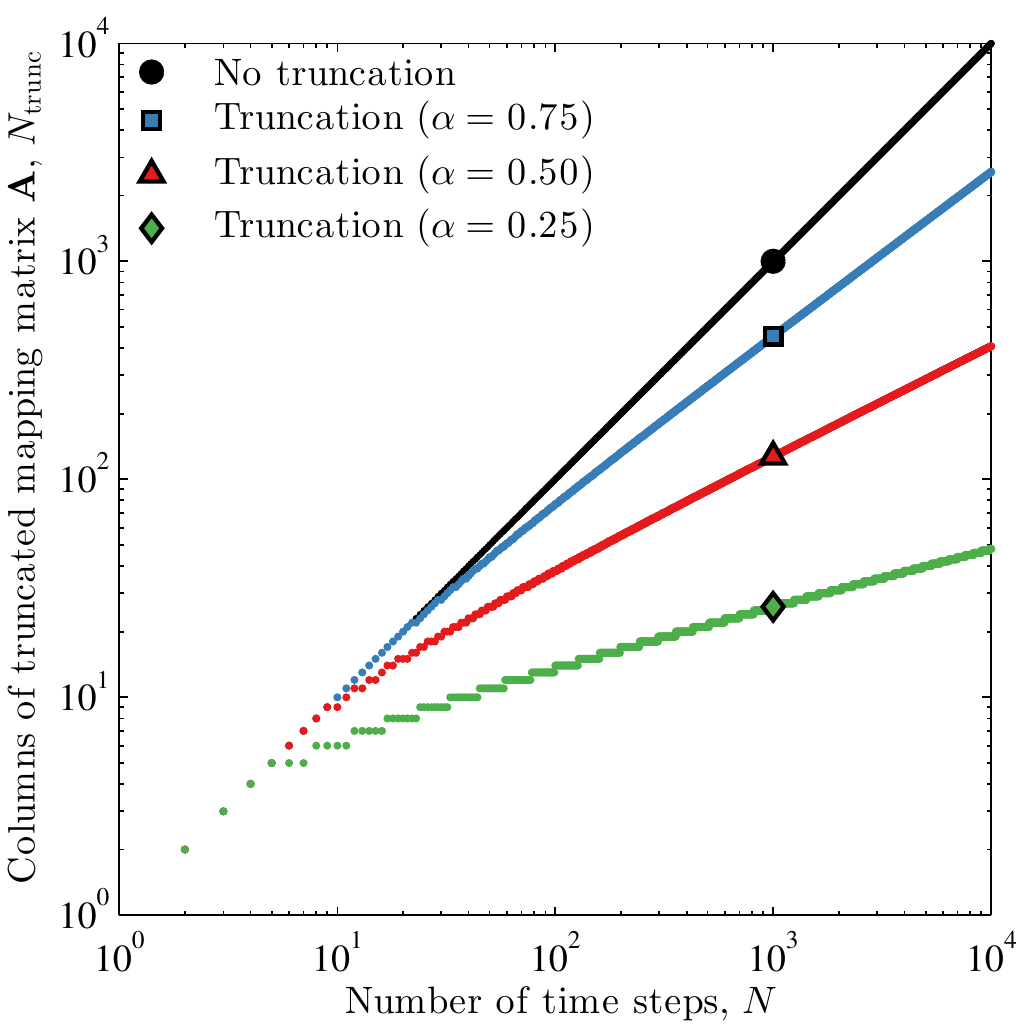}

\protect\caption{\label{fig:truncationComplexity}The number of columns in the mapping
matrix $\mathbf{A}$, truncated according to the inequality (\ref{eq:truncDefinition})
with a truncation tolerance of $\varepsilon=10^{-2}$. The gradient
of each case approaches $\alpha$ as the number of time steps $N$
grow large, suggesting the asymptotic form $N_{\mathrm{trunc}}\sim N^{\alpha}$.}
\end{figure}

The magnitude of the elements of the mapping matrix $\mathbf{A}$
is illustrated in Figure \ref{fig:truncationIllustration} for various
values of the fractional differentiation order $\alpha$. It can be
seen that, as $\alpha$ decreases, fewer elements are likely to contribute
to the solution mapping. This suggests that we can truncate the mapping
matrix at some point during its column-wise computation described
by Eq. (\ref{eq:convRecurrence}). Here, we will specifically consider
truncating the weighted sum (\ref{eq:singleMap}) corresponding to
the solution at the last time step. As a simplification, we will take
both integer and fractional order solutions to be constant and hence
equal, allowing us to remove all solution dependence and focus on
truncating the summation
\begin{equation}
\sum_{n>0}a_{Nn}=1.\label{eq:sumUnity}
\end{equation}
This expression can also be derived from the generating function representation
(\ref{eq:funRecurrence}) and is equivalent to stating that the rows
of the mapping matrix sum to unity. Now, by introducing a truncation
tolerance $0<\varepsilon<1$, which is proportional to the absolute
error incurred by the truncation, we can define the number of columns
in the truncated mapping matrix as the smallest integer $N_{\mathrm{trunc}}$
that satisfies
\begin{equation}
\sum_{n>N_{\mathrm{trunc}}}a_{Nn}\leq\varepsilon.\label{eq:truncOriginal}
\end{equation}
Evidently, to determine $N_{\mathrm{trunc}}$ using this inequality
requires computation of matrix elements that will ultimately be truncated.
Fortunately, using the row summation identity (\ref{eq:sumUnity}),
we can restate this inequality using known matrix elements
\begin{equation}
\sum_{1\leq n\leq N_{\mathrm{trunc}}}a_{Nn}\geq1-\varepsilon.\label{eq:truncDefinition}
\end{equation}
We can gain some insight into the asymptotic form of $N_{\mathrm{trunc}}$,
and hence any computational benefit of this truncation, by considering
the continuous analogue of this solution mapping, provided by Eq.
(\ref{eq:continuousMap}). As before, by choosing an integer order
solution that is constant, we find that
\begin{equation}
\int_{0}^{\infty}\mathrm{d}\tau A\left(\tau,t\right)=1,
\end{equation}
which is evident from the Laplace space representation (\ref{eq:kernelDef})
of $A\left(\tau,t\right)$ as being the normalisation condition for
an exponential distribution in $\tau$. By nondimensionalising in
terms of the finite difference time step indices, that is taking $t=N\Delta t$
and $\tau=n\Delta t^{\alpha}/w_{1}$, we find the continuous analogue
to the row summation identity (\ref{eq:sumUnity})
\begin{equation}
\int_{0}^{\infty}\mathrm{d}na_{Nn}=1,\quad a_{Nn}\equiv\frac{\Delta t^{\alpha}}{w_{1}}A\left(\frac{n\Delta t^{\alpha}}{w_{1}},N\Delta t\right),
\end{equation}
where both $n$ and $a_{Nn}$ are continuous here. Continuing with
the analogy, we can now choose to truncate this integral at the point
$n=N_{\mathrm{trunc}}$, resulting in the continuous analogue to truncation
tolerance definition (\ref{eq:truncOriginal})
\begin{equation}
\varepsilon\equiv\int_{N_{\mathrm{trunc}}}^{\infty}\mathrm{d}na_{Nn}=L_{\alpha}\sqrt[\alpha]{\frac{w_{1}N^{\alpha}}{N_{\mathrm{trunc}}}},
\end{equation}
where we have made use of the Lévy distribution representation (\ref{eq:kernelDef})
of the kernel $A\left(\tau,t\right)$. It is evident here that we
can make this truncation tolerance an arbitrarily small constant that
is independent of $N$ by choosing that $N_{\mathrm{trunc}}$ is directly
proportional to $N^{\alpha}$. As the discrete truncation tolerance
coincides with this continuous one in the limit of large $N$, we
should expect to find the asymptotic behaviour $N_{\mathrm{trunc}}\sim N^{\alpha}$
for the continuous case. Indeed, Figure \ref{fig:truncationComplexity}
shows precisely this as the size of the mapping matrix is increased
for select values of $\alpha$. Therefore, when truncated, an $N\times N$
mapping matrix becomes of size $N\times O\left(N^{\alpha}\right)$,
allowing for column-wise computation of it using the recurrence relationship
(\ref{eq:convRecurrence}) in only $O\left(N^{1+\alpha}\ln N\right)$.
Similarly, we can now find the fractional order solution at particular
instants in time in $O\left(N^{\alpha}\right)$. Finally, with this
truncation, it should be noted that we are no longer required to precompute
the mapping matrix in order to obtain a solution in a computational
complexity better than $O\left(N^{2}\right)$.

\section{\label{sec:Section-IV}Benchmark of the truncated mapping}

In this section, we will demonstrate the expected accuracy of the
truncated mapping solution described in Section \ref{sec:Section-III}
relative to the direct finite difference solution provided either
recursively or by the full mapping introduced in Section \ref{sec:Section-II}.
Specifically, we will consider the solution of the fractional relaxation
equation \citep{metzler2000random} 
\begin{equation}
_{0}^{\mathrm{C}}\mathcal{D}_{t}^{\frac{1}{2}}u\left(t\right)=u\left(t\right),\quad u\left(0\right)=1,\label{eq:fractionalODE}
\end{equation}
which we chose because it has the exact analytic solution \citep{haubold2011mittag}
\begin{equation}
u\left(t\right)=\mathrm{e}^{t}\left(1+\mathrm{erf}\sqrt{t}\right),\label{eq:odeAnalytic}
\end{equation}
where $\mathrm{erf}\left(x\right)\equiv2\pi^{-1/2}\int_{0}^{x}\mathrm{d}\xi\mathrm{e}^{-\xi^{2}}$
is the Gauss error function. Additionally, the finite difference solution
here can be found recursively by simply taking Eqs. (\ref{eq:fracRecurrence})
and (\ref{eq:intRecurrence}) with $\alpha=1/2$ and $\lambda=1$.

\begin{figure}
\includegraphics[scale=0.83]{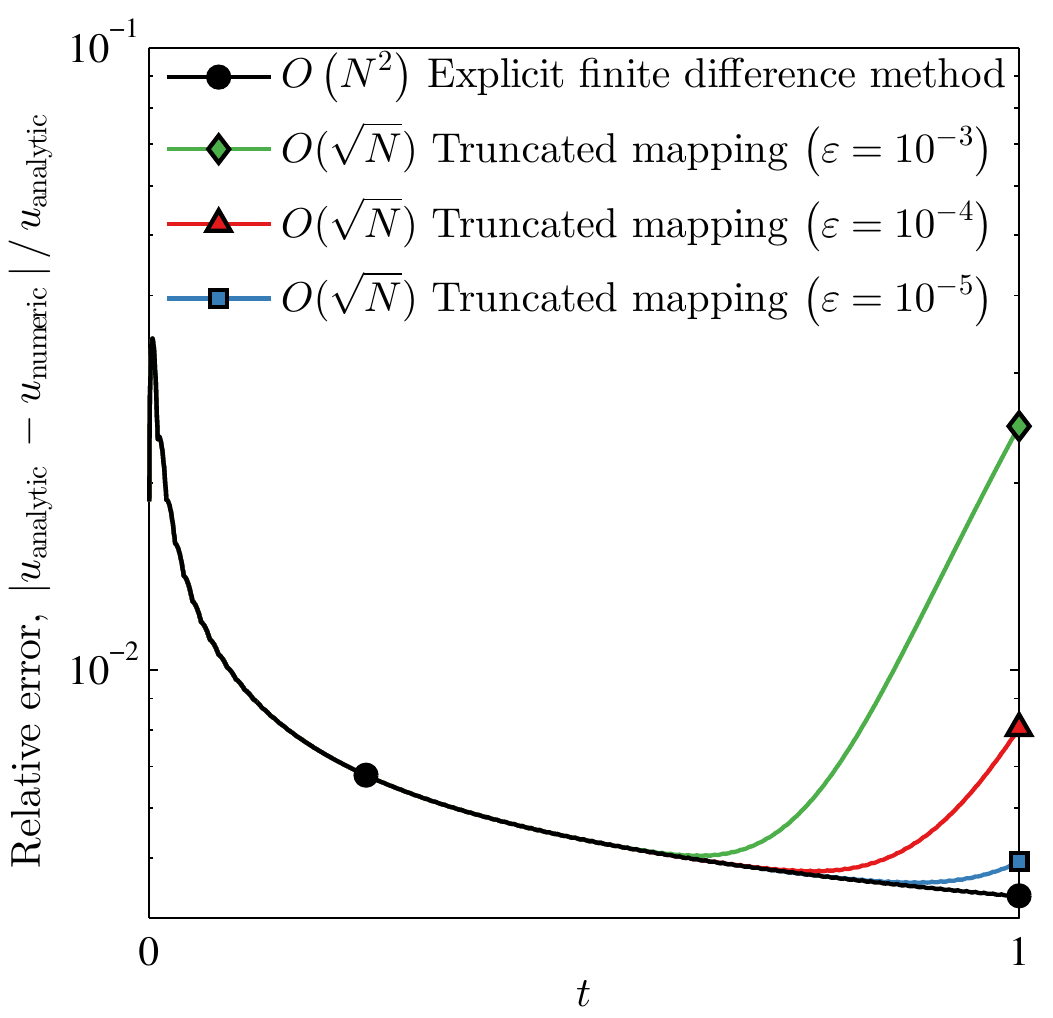}

\protect\caption{\label{fig:errorComparison}The error of an $N=100$ point finite
difference solution of the fractional relaxation equation (\ref{eq:fractionalODE})
relative to its analytic solution (\ref{eq:odeAnalytic}). The truncated
solution mapping described in Section \ref{sec:Section-III} is applied
for decreasing values of the truncation tolerance $\varepsilon$.
Note how the truncated mapping can be made to be arbitrarily accurate,
while still retaining its computational complexity of $O\left(\sqrt{N}\right)$.
The divergence in accuracy for late times stems from the truncation
of more terms at later time steps. To perform this plot, a truncated
mapping matrix $\mathbf{A}$ was precomputed in $O\left(N^{\frac{3}{2}}\ln N\right)$
and then truncated further as required.}
\end{figure}

Figure \ref{fig:errorComparison} shows that the truncated mapping
can be applied to find the solution to the fractional relaxation equation
(\ref{eq:fractionalODE}) to an accuracy comparable to the finite
difference method, while still maintaining an improved computational
complexity.

\section{\label{sec:Section-V}Application to the fitting of experimental
data}

Our approach is ideally suited to the acceleration of curve-fitting
problems where the solution defining the curve must be found repeatedly
and at relatively few points. In this section, we will demonstrate
this by fitting a fractional-order model to experimental data for
the current in a time-of-flight experiment for a disordered semiconductor.
As stated in Section \ref{sec:Section-I}, this can be described by
the fractional advection-diffusion model (\ref{eq:mainFDE}). This
model describes the charge carrier density in a thin sample held between
two large plane-parallel boundaries with all spatial variation occurring
normal to these boundaries. It will be assumed that the boundaries
are perfectly absorbing, providing the Dirichlet boundary conditions
\begin{equation}
u\left(t,0\right)=0=u\left(t,d\right),
\end{equation}
where $d$ is the thickness of the sample. We will choose the initial
distribution of charge carriers to be governed by the Beer-Lambert
law resulting in the exponential initial condition
\begin{equation}
u\left(0,x\right)\propto\mathrm{e}^{-ax},
\end{equation}
where $a$ is the absorption coefficient of the sample. We can use
the expression for the current in a time-of-flight experiment \citep{philippa2011analytic}
\begin{equation}
I\left(t\right)\propto\frac{\partial}{\partial t}\int_{0}^{d}\left(\frac{x}{d}-1\right)u\left(t,x\right)\mathrm{d}x,\label{eq:currentModel}
\end{equation}
to find the current directly from the number density solution of Eq.
(\ref{eq:mainFDE}). For spatial consideration, we will make use of
the centred finite difference approximations
\begin{eqnarray}
\frac{\partial}{\partial x}u\left(t,x\right) & = & \frac{u_{j+1}^{N}-u_{j-1}^{N}}{2\Delta x}+O\left(\Delta x^{2}\right),\\
\frac{\partial^{2}}{\partial x^{2}}u\left(t,x\right) & = & \frac{u_{j+1}^{N}-2u_{j}^{N}+u_{j-1}^{N}}{\Delta x^{2}}+O\left(\Delta x^{2}\right),
\end{eqnarray}
where $j=0,\dots,J$ is the spatial index, $J$ is the total number
of spatial nodes and subscripts have been used to denote spatial indexing
$u_{j}^{n}\equiv u\left(n\Delta t,j\Delta x\right)$. Hence, we can
enforce the boundary conditions by setting $u_{0}^{n}=0=u_{J}^{n}$
for all $n=0,\dots,N$. Applying these spatial derivative approximations,
in conjunction with Eq. (\ref{eq:fracDiscrete}) for approximating
the Caputo fractional derivative, results in the recurrence relationship
for the number density solution to Eq. (\ref{eq:mainFDE})
\begin{equation}
\mathbf{C}\mathbf{u}^{N}=\hat{w}_{N}\mathbf{u}^{0}+\sum_{n=1}^{N-1}\left(\hat{w}_{n}-\hat{w}_{n+1}\right)\mathbf{u}^{N-n},\label{eq:fadeRecurrence}
\end{equation}
where we have the tridiagonal matrix
\begin{equation}
\mathbf{C}\equiv\left[\begin{array}{ccc}
1-2r & r+s & 0\\
r-s & 1-2r & \ddots\\
0 & \ddots & \ddots
\end{array}\right],\quad r\equiv-\frac{D_{\mathrm{L}}\Delta t^{\alpha}}{w_{1}\Delta x^{2}},\quad s\equiv\frac{W\Delta t^{\alpha}}{2w_{1}\Delta x}.
\end{equation}

\begin{figure}
\includegraphics[scale=0.8]{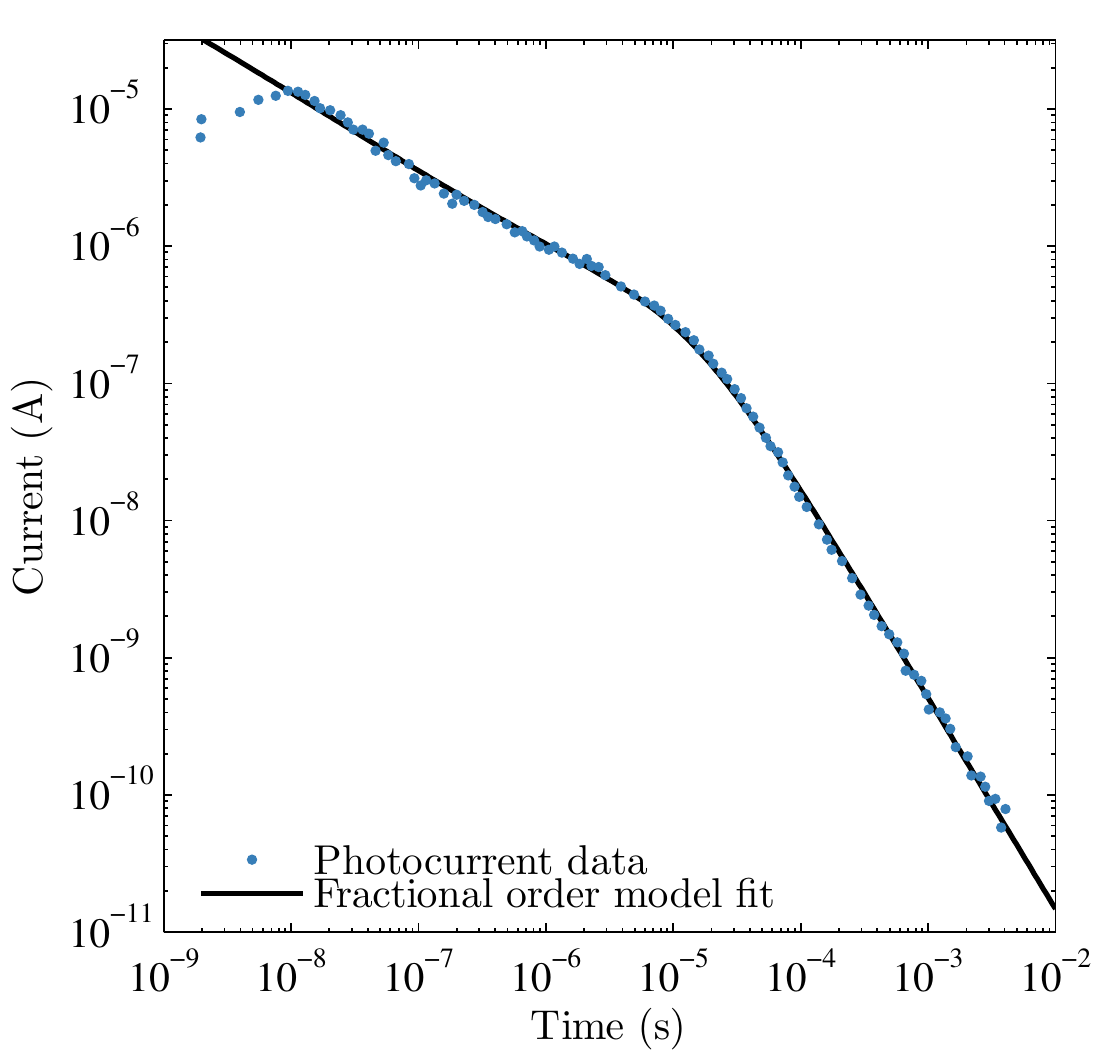}

\protect\caption{\label{fig:currentFit}A least squares fit of the model (\ref{eq:currentModel})
to the transient photocurrent in a sample of intrinsic hydrogenated
amorphous silicon a-Si:H at 160K (adapted from Ref. \citep{scher1991time}).
To within a confidence interval of $95\%$, the fitting algorithm
determined a severity of trapping of $\alpha=0.535\pm2\%$, a generalised
drift velocity of $Wd^{-1}t_{\mathrm{tr}}^{\alpha}=2.89\times10^{-1}\pm4\%$
and a generalised diffusion coefficient of $D_{\mathrm{L}}d^{-2}t_{\mathrm{tr}}^{\alpha}=6.07\times10^{-3}\pm21\%$,
where the ``transit time'' separating the current regimes has been
taken as $t_{\mathrm{tr}}\equiv10^{-5}\unit{s}$.}
\end{figure}

Figure \ref{fig:currentFit} plots photocurrent data alongside the
model (\ref{eq:currentModel}) fitted using a trust-region-reflective
non-linear least squares algorithm \citep{coleman1994convergence,coleman1996interior},
as implemented in the \texttt{lsqcurvefit} function \citep{lsqcurvefit}
located in MATLAB's Curve Fitting Toolbox.

To explore the computational benefits of applying the solution mapping
described in Section \ref{sec:Section-II} and its truncation described
in Section \ref{sec:Section-III}, we require the number density solution
when $\alpha=1$, corresponding to normal transport. Proceeding as
before, this time using Eq. (\ref{eq:intDiscrete}) for the approximation
of the first derivative, yields the recurrence relationship for the
integer order solution $v\left(t,x\right)$
\begin{equation}
\mathbf{C}\mathbf{v}^{N}=\mathbf{v}^{N-1}.
\end{equation}
As $\mathbf{C}$ is tridiagonal, we can step forward the fractional
order solution recurrence relationship (\ref{eq:fadeRecurrence})
in a time computational complexity of $O\left(J\right)$ \citep{press2007numerical}.
As such, the total computational complexity to determine the fractional
order solution in time and space becomes $O\left(N^{2}J\right)$.
Similarly, by applying the solution mapping we have a computational
complexity of $O\left(N^{2}J\ln N\right)$, which improves to $O\left(N^{1+\alpha}J\ln N\right)$
with truncation. The value of $\alpha$ present here can be estimated
by noting the asymptotic form of the current in a time-of-flight experiment
\citep{scher1975anomalous}
\begin{equation}
I\left(t\right)\sim\begin{cases}
t^{-\left(1-\alpha\right)}, & \mathrm{early\,times},\\
t^{-\left(1+\alpha\right)}, & \mathrm{late\,times},
\end{cases}\label{eq:currentAsymptote}
\end{equation}
which provides a criterion for recognising dispersive transport by
noting that the sum of the slopes of the asymptotic regions of a current
versus time plot on logarithmic axes is $-2$. In this particular
case, we can use this criterion to bound the severity of trapping
to within the interval $0.5<\alpha<0.55$.

\begin{figure}
\includegraphics[scale=0.82]{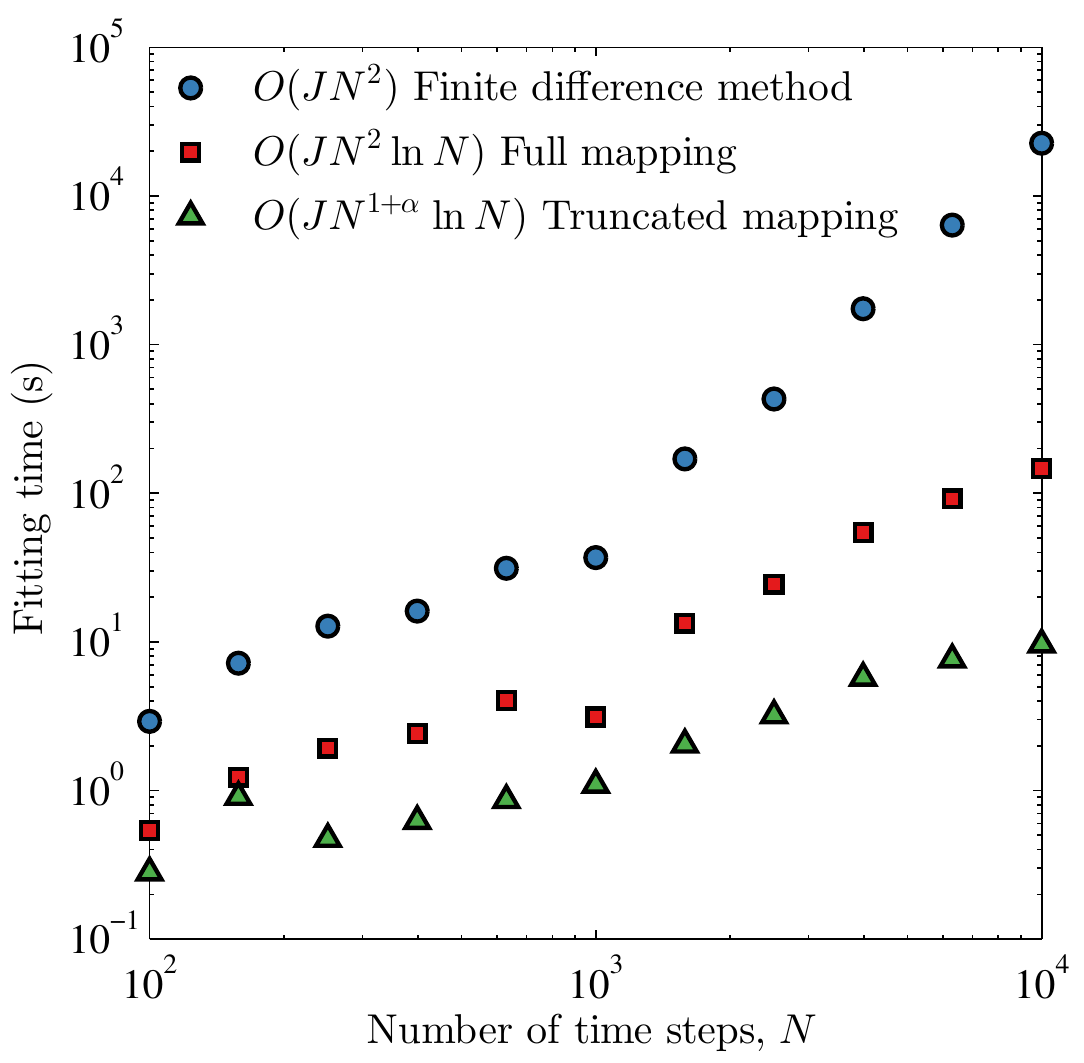}

\protect\caption{\label{fig:computationTime}Comparison of computation time versus
number of time steps for least squares fitting performed using the
finite difference method (\ref{eq:fadeRecurrence}), the accelerated
solution mapping developed here (\ref{eq:fullMap}) and the truncation
thereof defined by Eq. (\ref{eq:truncDefinition}). To maintain solution
accuracy, the truncation tolerance $\varepsilon$ was chosen to decrease
in proportion to $N$. It can be seen that the solution mapping without
truncation is two orders of magnitude faster than the recursive approach
for the largest problem size that was considered. With truncation,
this improves to a three orders of magnitude speed up.}
\end{figure}

Figure \ref{fig:computationTime} plots the computation time for fitting
the model (\ref{eq:currentModel}) to the photocurrent data considered
in Figure \ref{fig:currentFit} for an increasing number of time steps.
The observed fitting times do not increase monotonically with $N$.
This is due to the nature of the fast Fourier transform (FFT) algorithm.
The FFT is very sensitive to the prime factorisation of the input
size. For example, the FFT is fastest when $N$ is a power of 2, and
it is especially slow when $N$ is prime. Additional variations in
fitting time may be due to the curve fitting algorithm and the number
of iterations it requires to perform the fit.

\section{\label{sec:Section-VI}Concluding remarks and future work}

Finite difference solutions to fractional differential equations are
known to have a computation time that scales with the square of the
number of time steps. This stems mathematically from the global nature
of fractional differentiation, and physically can be interpreted as
a consideration of memory effects. In this study, we have related
the solution of the fractional diffusion equation (\ref{eq:mainFDE})
of order $0<\alpha<1$ to the solution of a the normal diffusion equation
(\ref{eq:mainIDE}) using a linear mapping in time Eq. (\ref{eq:fullMap}).
We have found that, for an $N$-point finite difference time discretisation,
we can use this mapping to improve upon the $O\left(N^{2}\right)$
time computational complexity of the finite difference method and
determine the solution at any instant in time in $O\left(N^{\alpha}\right)$,
given a precomputation of $O\left(N^{1+\alpha}\ln N\right)$. This
representation is especially useful in situations where the solution
must be found repeatedly, as then the relatively expensive precomputation
only has to be performed once. We have presented one such situation
in Section \ref{sec:Section-V} where we have successfully applied
this approach to fit the fractional advection-diffusion model (\ref{eq:mainFDE})
to experimental data for the current in a time-of-flight experiment.
For this we achieved computational speed ups in the range of one to
three orders of magnitude for the realistic problem sizes considered.

Although this work considered a fractional advection-diffusion model,
the mapping approach described in this paper is applicable for any
other linear spatial operator, including those of higher dimensions.
With modifications, this solution mapping can be generalised to consider
both the inclusion of a source term as well as higher order fractional
derivatives for which $\alpha>1$.
\begin{acknowledgments}
\appendix
We gratefully acknowledge the funding of the Australian Research Council
(Discovery and Centres of Excellence programs) and the Queensland
Smart Futures Fund.
\end{acknowledgments}

\bibliography{references}

\end{document}